\documentclass[]{spie}  %>>> use for US letter paper
\usepackage[dvipsnames]{xcolor}
\usepackage{amssymb}
\usepackage{mathrsfs}
\usepackage{amsmath}
\usepackage{subfigure}
\usepackage{url}
\usepackage{stackrel} 
\usepackage{courier}
\usepackage[all,cmtip]{xy} 
\usepackage{multicol} 
\usepackage{verbatim} %  useful for \begin{comment} \end{comment}  
\usepackage{mathtools}  
\usepackage[colorlinks=true, allcolors=blue]{hyperref}
\usepackage{braket}
\usepackage{breqn}
\usepackage{multicol}  
\usepackage{enumerate} 
\usepackage{bookmark} 
\usepackage{ytableau}
\usepackage{microtype}
\usepackage{float,titlesec}
 
\usepackage{algorithm}
\usepackage[noend]{algpseudocode}

\catcode`~=11 % make LaTeX treat tilde (~) like a normal character
\newcommand{\urltilde}{\kern -.15em\lower .7ex\hbox{~}\kern .04em}  
\catcode`~=13 % revert back to treating tilde (~) as an active character

\makeatletter
\newif\if@borderstar
   \def\bordermatrix{\@ifnextchar*{%
       \@borderstartrue\@bordermatrix@i}{\@borderstarfalse\@bordermatrix@i*}%
   }
   \def\@bordermatrix@i*{\@ifnextchar[{\@bordermatrix@ii}{\@bordermatrix@ii[()]}}
   \def\@bordermatrix@ii[#1]#2{%
   \begingroup
     \m@th\@tempdima8.75\p@\setbox\z@\vbox{%
       \def\cr{\crcr\noalign{\kern 2\p@\global\let\cr\endline }}%
       \ialign {$##$\hfil\kern 2\p@\kern\@tempdima & \thinspace %
       \hfil $##$\hfil && \quad\hfil $##$\hfil\crcr\omit\strut %
       \hfil\crcr\noalign{\kern -\baselineskip}#2\crcr\omit %
       \strut\cr}}%
     \setbox\tw@\vbox{\unvcopy\z@\global\setbox\@ne\lastbox}%
     \setbox\tw@\hbox{\unhbox\@ne\unskip\global\setbox\@ne\lastbox}%
     \setbox\tw@\hbox{%
       $\kern\wd\@ne\kern -\@tempdima\left\@firstoftwo#1%
         \if@borderstar\kern2pt\else\kern -\wd\@ne\fi%
       \global\setbox\@ne\vbox{\box\@ne\if@borderstar\else\kern 2\p@\fi}%
       \vcenter{\if@borderstar\else\kern -\ht\@ne\fi%
         \unvbox\z@\kern-\if@borderstar2\fi\baselineskip}%
         \if@borderstar\kern-2\@tempdima\kern2\p@\else\,\fi\right\@secondoftwo#1 $%
     }\null \;\vbox{\kern\ht\@ne\box\tw@}%
   \endgroup
   }
\makeatother

\titleformat{\paragraph}
{\normalfont\normalsize\bfseries}{\theparagraph}{1em}{}
\titlespacing*{\paragraph}
{0pt}{3.25ex plus 1ex minus .2ex}{1.5ex plus .2ex}

\allowdisplaybreaks

\title{Monte Carlo methods on a fixed volume system of Silicon-Germanium atoms}  
\author{Mee Seong Im, Ph.D.} 

\affil{Department of Mathematical Sciences, U.S. Military Academy, West Point, NY 10996}
\authorinfo{Author information: meeseongim@gmail.com}  
\begin{document}
\maketitle  
  
\begin{abstract}   
Since the electrons of a silicon-germanium system are bounded, external quantum effects are negligible. 
In this manuscript, we hold the volume constant while varying all other parameters, such as pressure, temperature, germanium chemical potential (or germanium concentration), energy, mole number and atomic bond structure, resulting in an observation of hysteresis in the system.
\end{abstract}   

\keywords{Silicon, germanium, Monte Carlo method, hysteresis, fixed volume system}

\setcounter{tocdepth}{1}

\section{Introduction}
\label{section:introduction}  
 
One way to detect whether the computer simulator is biased is by comparing a simulated output with its mathematical calculations. If the simulated output is different from its expectation, then the simulated model is incorrect. Thus Monte Carlo (MC) techniques have been developed when certain random number generators were not shown to be purely random\cite{binder1993monte,landau2014guide}. 
Today, MC randomness is used to predict the probability in stock market and economy growth, equity index annuities, consumer statistics, national survey, etc\cite{hammersley2013monte,hastings1970monte,kalos2009monte,smith2013sequential}. In this paper, two species of atoms, closely related but with different atomic sizes, were used and studied using the MC method. 
The set of atoms is in a closed system, free from external forces, energy, and pressure. Thus we will assume that the entire system, consisting of pressure, wind, gravity, heat, and particles in air, external to the tested system has negligible affect.

Let $L$ denote as the length of one unit cell. So the volume of the investigated system is $L^3$. 
Since there are eight atoms per unit cell, there are total of $8 \times L^3$ atoms per length $L$ system. 
We investigated a system of three various sizes: $L$ $=$ $4$, $6$, and $8$, which has $512$, $1728$, and $4096$ atoms, respectively. The $L = 8$ system is still very small, for it is equivalent to $6.80173 \times 10^{-21}$ moles, and there are $6.022 \times 10^{23}$ atoms/mole. So using computer simulations, our conclusions are a result from a specific set of boundary conditions. In addition, a system of $4096$ atoms in a MC program takes approximately thirty six to forty hours to execute $2.1$ million runs. Though we started computing the data every $50$ MC steps,  
we gathered data every $300$ MC steps near the end of our simulation in order to observe stability in the system. 

Although the size of the output files became smaller, total computation time did not reduce according to the computing steps. For the $512$ atom system, we executed $600,000$ to $900,000$ MC runs, and $900,000$ MC runs for the system with $1728$ atoms. For each execution, we ran sufficiently enough runs to achieve our desired state. This was determined both interchangeably by comparing with the other data and independently with various plotting programs. There was one exception, where $2.1$ million runs were not enough for a $4096$ atom system. This is further discussed in \S\ref{section:hysteresis}. 

During this execution, the MC moves, or replaces, one atom at a time to reduce the system's total energy, which the process is called \textit{semi-grand-canonical ensemble}\cite{heermann1990computer,adler2002recent}. In the semi-grand-canonical ensemble, the total number of the system's atoms is held constant, but the number of the species may vary. Thus the number of each species may change, not the total number of atoms in the system. Lower energy for a system means that atoms are in a less excited state; more of the atoms' electrons are in ground state, leading to a state of order. For example, if we heat up a pot of ice, it melts into liquid. Then, as the ice temperature inside the pot increases, the ice, or now water, becomes gas. Investigating from a chemist's point of view, the water molecule $H_2O$ separates into $H_2$ and $O_2$ gas. From the physics perspective, the increased heat induces the electrons move to their excited state, breaking shared bonds, and allowing the atoms to move freely. So with less energy, there are less active electrons and less entropy. A state of symmetry develops, which is a state when distinct geometries become vital in the system. 

In this paper, a system of silicon (Si) and germanium (Ge) atoms were investigated. 
Since no electron clouds surround these atoms, the system behaves as an insulator. 
Furthermore, 
since quantum effects may be eliminated, computation, simulation and analysis are greatly simplified. 

In previous simulations, pressure was held fixed while other extensive parameters, such as volume, temperature, germanium chemical potential or germanium concentration, energy, mole number and atomic bond structure, were varied.  
Fixing the volume of the system instead of the pressure and by varying the germanium chemical potential and the temperature of the system, surprising results were revealed in its ordered state.

\section{Properties of Silicon-Germanium atoms} 
\label{section:properties-Si-Ge}
Silicon (Si) and germanium (Ge) atoms have $p$-block configuration. That is, silicon valence configuration is $1s^2 2s^2 2p^6 3s^2 3p^2$, while the germanium valence configuration is $1s^2 2s^2 2p^6 3s^2 3p^6 3d^{10} 4s^2 4p^2$. The atomic weight for Si is $28.08553$, while it is $72.641$ for Ge. Si is dark grey with a bluish tinge, while Ge is grayish white. Both are solid at $298 K$, semi-metallic, and have diamond crystal lattice\cite{conwell1952properties,conwell1958properties,shiraki2011silicon,kelires1996microstructural,lannoo1991atomic,lannoo2013atomic}. 

Silicon is abundant in the universe, e.g., the sun. About $25.7\%$ of the earth's crust, by weight, is silicon, with it being the second most abundant element. It is often found as sand, quartz, rock crystal, amethyst, agate, flint, jasper and opal. It's also found in asbestos, feldspar, clay and mica. Large portion of animal and plant life depend on Si. For example, silicon is extracted from both fresh and salt water to grow and to nurture cell walls. It is also an important component in metals, such as steel. It serves as abrasives, conducts electricity, transports information, and holds large amounts of data in computer chips. 

Silicon can easily be isolated using silica ($\text{SiO}_2$) and graphite, or silicon chloride ($\text{SiCl}_4$) and hydrogen. The resulting molecules are pure silicon and two carbon monoxides for the former, and silicon and four hydrochlorides for the latter. 

Germanium is a rare element since it is not abundant. It is however an important semiconductor. Germanium plays an important role in solid state electronics. When combined with other elements, it has high ductility, chemical resistivity, infrared transmission and high refractive index. It is a grayish white metal which is crystal-like and brittle in its pure state. It is used in diodes, transistors, and light and temperature sensors. Some germanium compounds have other uses, such as killing certain bacteria. It has also shown to ``boost human immune system, normalize high blood pressure and cholesterol, protect the body against harmful cellular aberrations and abuse, provide some pain relief, alleviate rheumatoid arthritis symptoms and generally normalize physiological functions"\cite{Grace}. It also raises the cell's supply of oxygen, which is vital to the standards of healthy living.  

Similarly, germanium can also be isolated using germanium dioxide ($\text{GeO}_2$), with carbon or hydrogen, or germanium chloride ($\text{GeCl}_4$) and hydrogen. Its byproduct is germanium with carbon monoxide or water, or germanium and hydrochloride. 

In this manuscript, the primary difference between Si and Ge that we consider is the size difference by  $4\%$. Germanium is about $4\%$ larger than silicon. In addition, we will mainly focus on $L = 8$ system. As for $L$ $=$ $4$ and $6$ systems, they will be used to compare system dependences if any exist. Since systematic patterns can easily be seen in the bigger system, and since bigger systems correspond better to universal generalization, $L = 8$ system will also be discussed.

\section{Fixed volume system}
\label{section:fixed-volume-system}

We denote the $512$, $1728$, and $4096$ atom system as $4$, $6$, $8$-system, respectively. For the $8$-system, temperatures ranged from $0.00150$ to $0.00500$ eV were investigated. Tavazza\cite{Tavazza} investigated a similar problem, but only with the volume of the system at $1.03$ and at $L = 10$ and $12$. Since Ge is $4\%$ bigger than Si, Tavazza fixed the volume at $3\%$, where Ge were distributed in a smaller volume system. Thus, Ge-Ge repulsion occurred since germaniums want to be in their normal state. As for the silicon atoms, they were forced to be further apart. This created a strong attraction among the silicon atoms. 

By her previous simulations, the range of temperatures for the system with volume at $1.02$ and at $L = 4$, $6$, and $8$ that will be vital in investigating phase transition, from order to disorder, hysteresis, size dependence properties, Si and Ge behavior, etc. were estimated. Her range of temperatures were slightly higher than $L = 8$ temperatures. At $70\%$ concentration of germanium atoms, the phase transition (order to disorder border line) temperature is $0.00415$ eV for $L = 12$ and $0.00390$ eV for $L = 8$. On average, the phase transition temperature for Tavazza's $12$-system is about $6.0\%$ higher than the $8$-system. The $8$-system temperature ranged from $0.00125$ eV to $0.00500$ eV, where increments of $0.00005$ to $0.00025$ were simulated. In addition, the range of germanium chemical potential that corresponds to germanium concentration for the $12$-system is slightly higher than for the $8$-system. The germanium chemical potential for the $8$-system ranged from $0.001433$ eV to $0.001505$ eV, where the average increment increased by $0.000021$ eV, where $L = 12$ has a volume of $1.03$, while $L = 8$ has a volume of $1.02$. 

The Ge-Si system volume at $1.03$ meant stronger attraction among the Si atoms than repulsion from Ge. In addition, the silicon atoms are lighter than germaniums. Thus, silicon atoms relatively shift even further than germaniums, although their forces are not as strong unless they were the heavier atoms. As for $1.02$ volume, there is equal and opposite amount of attraction and repulsion by Si and Ge, respectively. But the silicon atoms are the ones to shift the system, more often than Ge because of the weight of each species.

 \begin{figure}[htbp]
\centering
\subfigure[]
{\includegraphics[width=5.15cm]{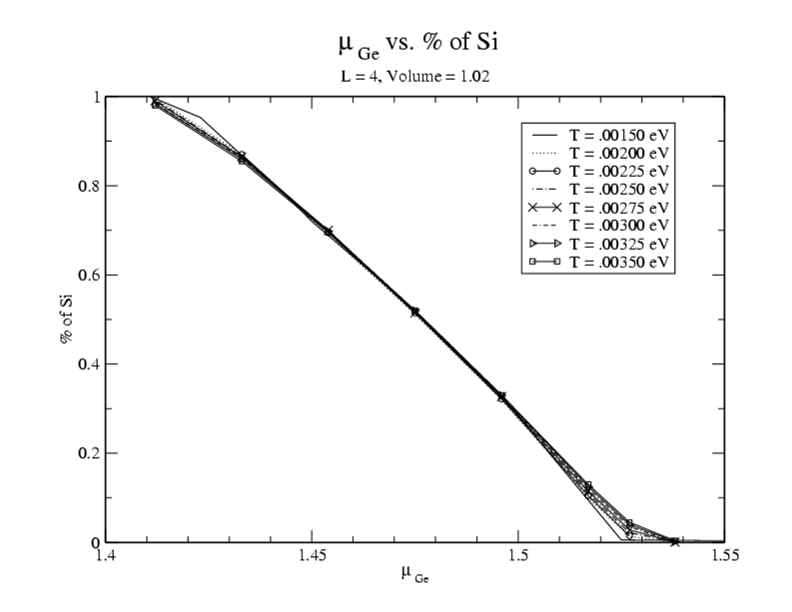}}\qquad 
\subfigure[]
{\includegraphics[width=5.15cm]{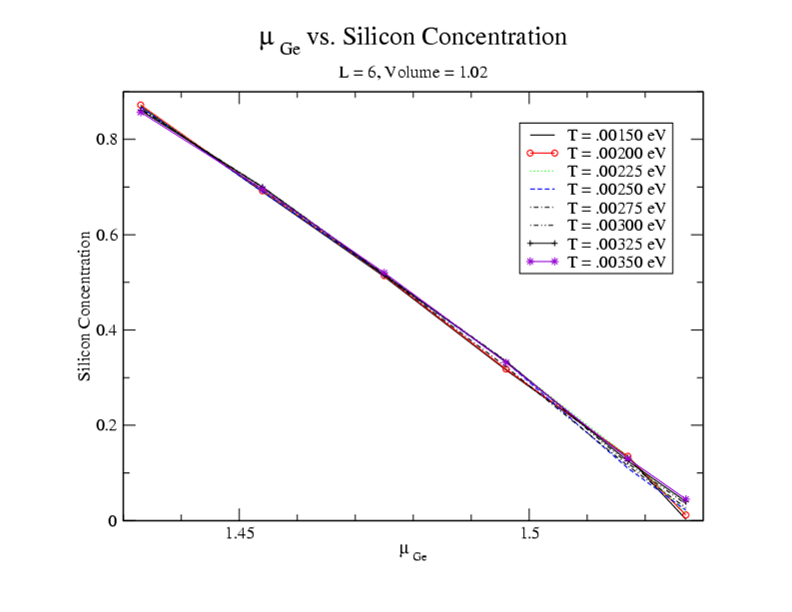}} \qquad  
\subfigure[]
{\includegraphics[width=5.15cm]{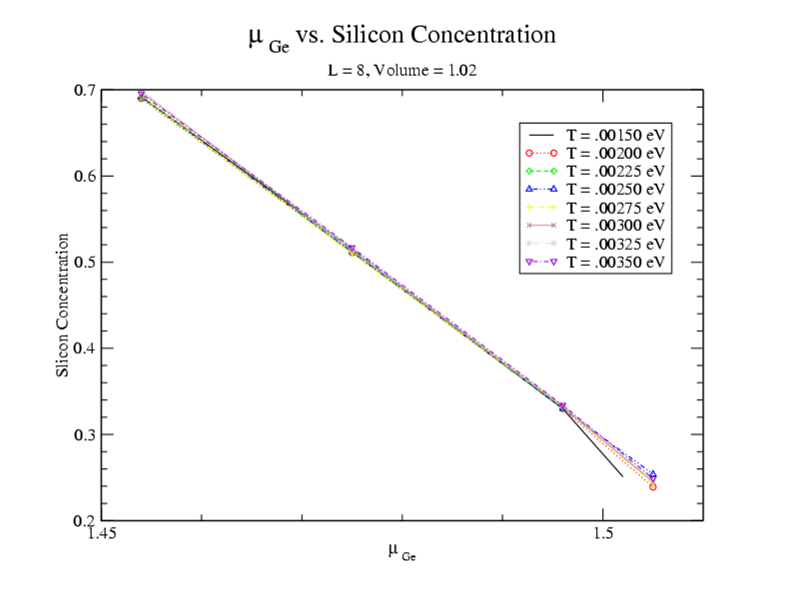}}
\label{fig:si-ge-01}
\end{figure}

\section{Monte Carlo simulation in statistical physics}
\label{section:Monte-Carlo-simulation}  

The total energy for each different system size is not dependent on the volume. Silicon concentration, and thus germanium concentration, does not instantly depend on the changes of the system size. However, at the extremes of germanium concentration, i.e., near the high or the low germanium concentration end, germanium concentration is slightly higher as the system size increased. Hysteresis, a retardation of an effect when the forces acting upon a body are changed, is predicted to affect these changes\cite{choi2002observation,hu1979superconducting,pharr1992electrical,yamazaki1975semiconductor}.  

Not surprisingly, for any fixed germanium chemical potential, the average number of silicon-to-silicon bonds is higher for a bigger system. And by symmetry, for any germanium chemical potential, the average number of germanium-to-germanium bonds is higher for each increasing system size. For any chosen germanium chemical potential, the change in energy for the temperature ranges of $0.00150$ eV to $0.00350$ eV is $0.0027$ eV for $L = 4$, $0.0034$ eV for $L = 6$, and $0.00405$ eV for $L = 8$. 

For any germanium chemical potential, energy increases evenly as temperature increases for any system size. And as temperature increases from $0.00150$ eV to $0.00350$ eV, the average number of silicon-to-silicon bonds is broken more quickly. Similarly, the average number of germanium-to-germanium bonds break with the raising of the system temperature. As for the silicon concentration, there appears to be hysteresis, as the system temperature is varied.  

\texttt{aviz} is a graphing program where one may plot actual visualization of the atomic system. It takes a momentary shot of the system at a particular time, and plots all the atoms. In our cases, it's the moment the executed file was completed. Note that left surface of the system, or box, is connected to the right side, and the top surface is connected to the bottom of the box. As seen in any introductory chemistry course, silicon atoms tend to bond with other silicon, while germanium tends towards other germanium atoms. One expects all the silicon atoms to be in a spherical shape for equilibrium and to maximize entropy (and minimize energy), or one size of the system to have solely the silicon atoms while the other side of the system to contain only germanium. However, none of these hypotheses were the results. After giving each run a sufficient time to reach equilibrium, silicon planes were formed, and surrounding the planes were Ge. And as germanium concentration decreased, or the number of the silicon atoms in the system increased, another plane formed perpendicular to the current plane. The third formed silicon plane was perpendicular to both of the existing silicon planes. As germanium concentration continued to decrease, a silicon plane, parallel to the first plane but perpendicular to the other two planes, developed. As for Tavazza's $10$ and $12$-system at volume $1.03$, all the silicon planes have similar thickness: each silicon plane has no more than a thickness of two or three layers of silicon atoms. As for the $8$-system at $1.02$ volume, mutually perpendicular silicon planes, parallel to the regular rectangular coordinate axes, always formed. However, some planes were thicker than others, while other silicon planes appeared to simply take over at least 2/3 of that axis in that particular direction. In all cases, the silicon atoms always formed a plane, rather than any other geometric shape.

\begin{figure}[htbp]
\centering
\begin{minipage}[b]{.4\textwidth}
\includegraphics[width=5cm]{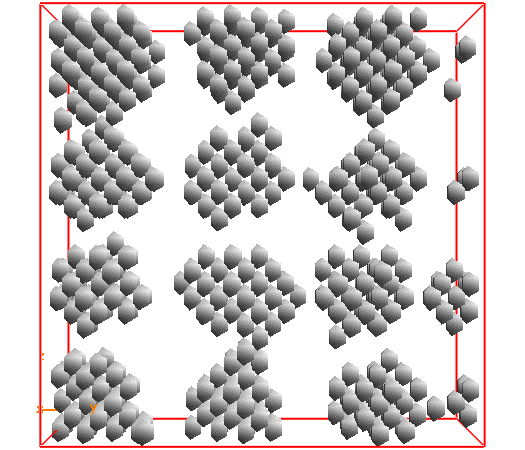}
\caption{Volume = 1.02, $L = 8$. 
Germanium chemical potential 0.001433 eV, temperature 0.00125 eV. 
Germanium shown. Seven silicon planes. 2.1 million MC runs. Front view.}
\label{fig:aviz-01}
\end{minipage} \qquad 
\begin{minipage}[b]{.4\textwidth}
\includegraphics[width=5cm]{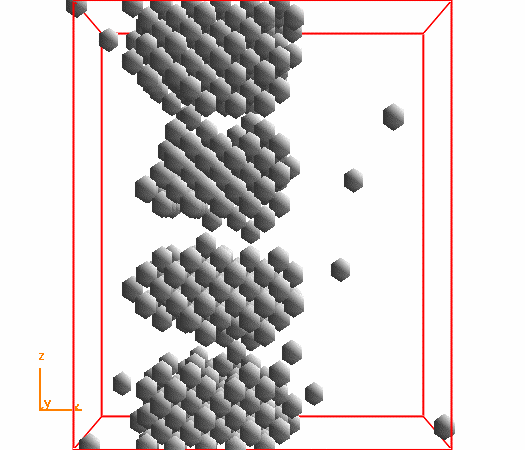} 
\caption{Volume = 1.02, $L = 8$. 
Germanium chemical potential 0.001433 eV, temperature 0.00125 eV. 
Germanium shown. Seven silicon planes. 2.1 million MC runs. $90^{\circ}$ rotated view.}
\label{fig:aviz-01}
\end{minipage}
\end{figure}

\begin{figure}[htbp]
\centering
\begin{minipage}[b]{.4\textwidth}
\includegraphics[width=6cm]{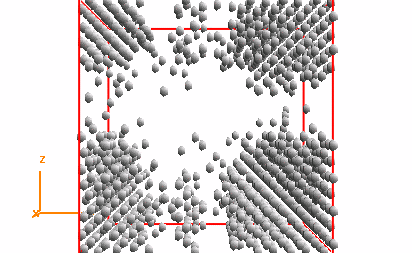} 
\caption{Germanium chemical potential of 0.001454 eV, temperature 0.00250 eV. 
Ge shown. Five silicon planes, two planes along the vertical axis. 
Executed over 2.1 million MC runs.} 
\label{fig:aviz-03}
\end{minipage} \qquad  
\begin{minipage}[b]{.4\textwidth}
\includegraphics[width=6cm]{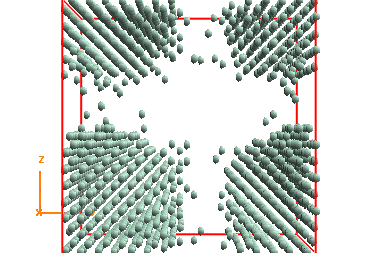}
\caption{Germanium chemical potential of 0.001475 eV, temperature 0.00275 eV. 
Ge shown. Two silicon planes. 
Executed over 2.1 million MC runs.}
\label{fig:aviz-04}
\end{minipage}
\end{figure}

\begin{figure}[htbp]
\centering
\begin{minipage}[b]{.4\textwidth}
\includegraphics[width=6cm]{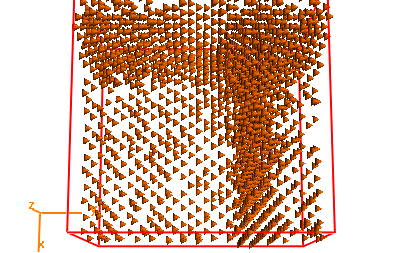}
\caption{Germanium chemical potential 0.001485 eV, temperature 0.00300 eV. 
Silicon shown. Three silicon planes. 2.1 million MC runs.}
\label{fig:aviz-05}
\end{minipage}\qquad 
\begin{minipage}[b]{.4\textwidth}
\includegraphics[width=6cm]{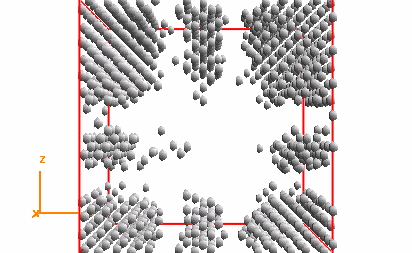} 
\caption{Germanium concentration increasing and all else held fixed.
Volume = 1.02, $L = 8$.
Temperature = 0.00200 eV. 2.1 million MC runs. 
Germanium chemical potential 0.001454 eV. Germanium shown. Five silicon planes, where one is parallel to the image.}
\label{fig:aviz-06}
\end{minipage}
\end{figure}

\begin{figure}[htbp]
\centering
\begin{minipage}[b]{.4\textwidth}
\includegraphics[width=6cm]{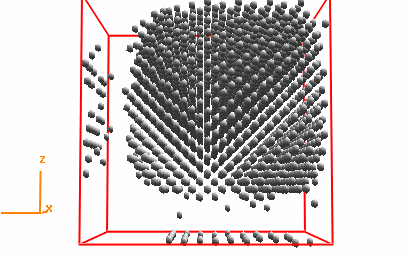}
\caption{Germanium concentration increasing and all else held fixed.
Volume = 1.02, $L = 8$.
Temperature = 0.00200 eV. 2.1 million MC runs. 
Ge chemical potential 0.001475 eV. Germanium shown. Two silicon planes.}
\label{fig:aviz-07}
\end{minipage} \qquad 
\begin{minipage}[b]{.4\textwidth}
\includegraphics[width=6cm]{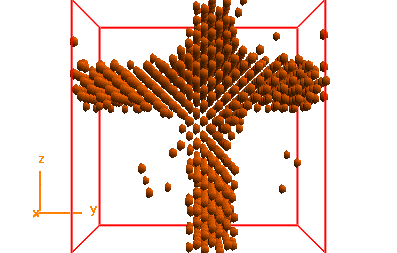} 
\caption{Germanium concentration increasing and all else held fixed.
Volume = 1.02, $L = 8$.
$T = 0.00200$ eV. 2.1 million MC runs. 
Ge chemical potential 0.001496 eV. 
Silicon shown. Two silicon planes.}
\label{fig:aviz-08}
\end{minipage} 
\end{figure}

\begin{figure}[htbp]
\centering
\subfigure[Germanium concentration increasing and all else held fixed.
Volume = 1.02, $L = 8$.
Temperature = 0.00200 eV. 2.1 million MC runs. 
Germanium chemical potential 0.001505 eV. 
Silicon shown. One silicon plane.]
{\includegraphics[width=5cm]{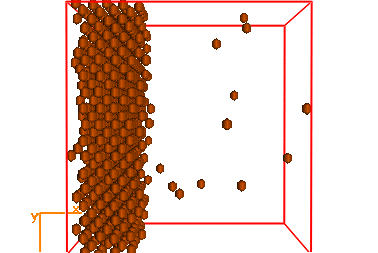}} \qquad 
\subfigure[Germanium concentration increasing and all else held fixed.
Volume = 1.02, $L = 8$.
$T= 0.00150$ eV. 2.1 million MC runs. 
Ge chemical potential 0.001433 eV. Germanium shown. Six silicon planes. Front shot.]
{\includegraphics[width=4cm]{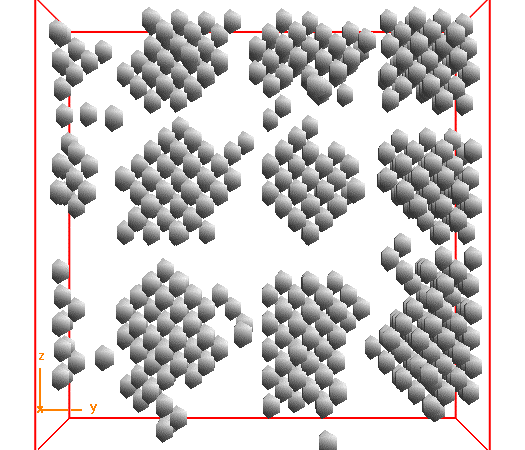}}\qquad 
\subfigure[Germanium concentration increasing and all else held fixed.
Volume = 1.02, $L = 8$.
$T = 0.00150$ eV. 2.1 million MC runs. 
Ge chemical potential 0.001433 eV. Germanium shown. Six silicon planes. Side shot, rotated $90^{\circ}$ clockwise.]
{\includegraphics[width=4cm]{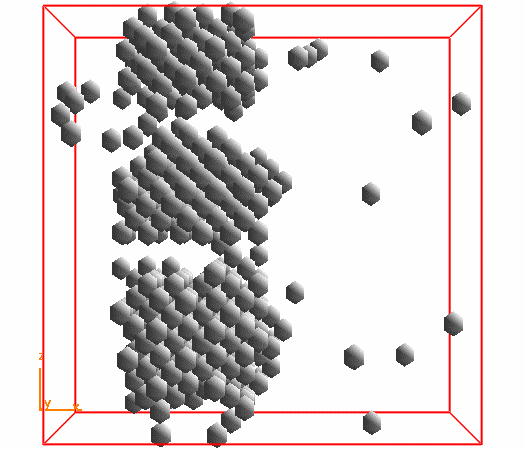}}
\label{fig:aviz-09}
\end{figure}

\begin{figure}[htbp]
\centering
\subfigure[Germanium concentration increasing and all else held fixed.
Volume = 1.02, $L = 8$.
Temperature = 0.00150 eV. 2.1 million MC runs. 
Germanium chemical potential 0.001445 eV. Germanium shown. Five silicon planes, one parallel to the $y$-axis. ]
{\includegraphics[width=6cm]{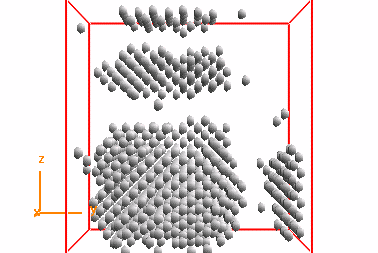}}\qquad 
\subfigure[Germanium concentration increasing and all else held fixed.
Volume = 1.02, $L = 8$.
Temperature = 0.00150 eV. 2.1 million MC runs. 
Germanium chemical potential 0.001454 eV. Germanium shown. Four silicon planes, one parallel to the $x$-axis. ]
{\includegraphics[width=6cm]{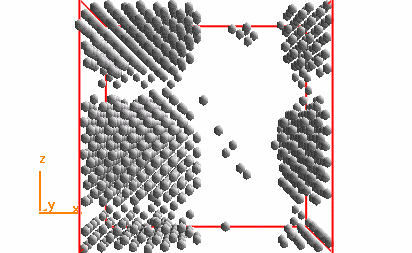}}
\label{fig:aviz-12}
\vspace{-3.5mm}
\end{figure}

\begin{figure}[htbp]
\centering
\subfigure[Germanium concentration increasing and all else held fixed.
Volume = 1.02, $L = 8$.
Temperature = 0.00150 eV. 2.1 million MC runs. 
Germanium chemical potential 0.001475 eV. Germanium shown. Four silicon planes. Front view.]
{\includegraphics[width=6cm]{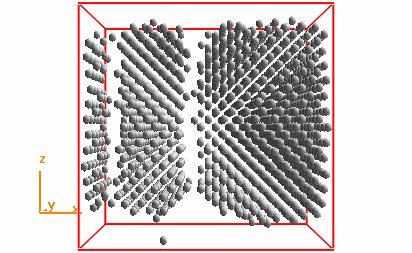}} \qquad 
\subfigure[Germanium concentration increasing and all else held fixed.
Volume = 1.02, $L = 8$.
Temperature = 0.00150 eV. 2.1 million MC runs. 
Germanium chemical potential 0.001475 eV. Germanium shown. Four silicon planes. Rotated view, rotated $90^{\circ}$ clockwise.]
{\includegraphics[width=6cm]{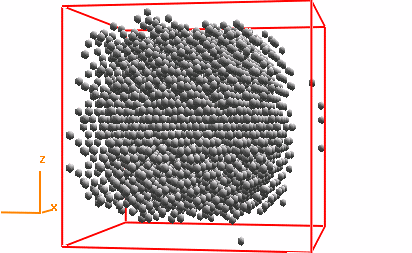}}
\label{fig:aviz-14}
\end{figure}

Some reasons that could explain the formation of the silicon planes are the difference in silicon and germanium sizes and the constant volume assumption for each system. Since germanium is $4\%$ bigger than silicon, all the Si-Ge bond for a particular system is not equivalent, since some Si-Ge bond will be farther apart than another set of silicon-to-germanium bond. Thus, there are greater bond tensions among some bonds than others. 

In Tavazza's observation, silicon atoms have a great tension to be closer to each other than Ge-Ge bonds. So the bond separation among the silicon atoms creates greater stress than the germanium bonds, although the germaniums have more mass. In the $8$-system with $1.02$ volume case, both the silicon atoms and the germaniums have equal and opposite forces to attract and to repel. So as the silicon concentration increases, planes form without affecting its thickness. The mass of the species shows which specie is more likely to move and adjust for equilibrium. Also in the $8$-system, whether the thick silicon planes were originally several planes combined as one and formed at different times or one thick plane that was formed all at once, there appears to be no parameter on the thickness of the silicon planes. 

When considering applications for this system, it depends on the germanium concentration and for what it is being used. Scientists, engineers, and doctors have different uses, for laboratory experiments, technology, or medicine, but the silicon behavior does not heavily affect their usage, unless the composition of the object in consideration has a significant amount of silicon and germanium. 

Ge-Si phase transition was determined in the following two ways. First, choose a germanium chemical potential for a specific system at a low enough temperature such that its final state is in an ordered state. Then with all conditions fixed, increase the temperature slowly. As soon as the system passes through its phase transition line, there will be disorder. That is, there will be a sudden increase in energy or a rapid decrease in the average Si-Si bonds and the average Ge-Ge bonds since the electrons for each atom will rise to its excited state, leading to an increase in their velocities and disorder. So, when plotting results from low temperature and the increased temperatures together, e.g., number of MC steps verses average number of Ge-Ge bonds, there is a gap between ordered states and disordered states. The gap is considered as an error distance which will be replotted as an error bar. If considering the plane as MC steps verses Ge-Ge bonds, the phase transition is determined by subtracting the highest disordered state from the lowest ordered state. Each error bar is reduced as the execution is repeated with its temperature set within its error bars. The range of guessing the exact phase transition temperature is reduced. Another way to determine phase transition is by first, plotting the visualization of each system. For each germanium chemical potential, execute at different temperatures, and then count the number of silicon planes. If no distinct silicon planes can be seen, distinguish these systems as disordered. Confusion may occur if the MC runs were not long enough, thus leading to a guess of an ordered state as disordered, and vice versa. Plot this data on a graph as Ge chemical potential verses temperature, or Ge concentration verses temperature. Using either of the two methods, a similar graph is obtained. 
 
\section{Hysteresis}
\label{section:hysteresis}

Sometimes, as the system's energy decreases during a particular MC run, the system can get stuck in a state called \textit{metastable state}. Metastable state is a state that appears to be in final equilibrium because the system appears organized, i.e., distinct silicon planes and small variations of energy range for a long time. But this is a state where the system is stuck at a particular temperature until the MC method changes an appropriate number of Si for Ge, and vice versa, causing a drop in the system's total energy. For the system with germanium chemical potential of $1.496\times 10^{-3}$ eV at temperature $0.00275$ eV, one expected for the system to be in its final equilibrium at two million MC steps. However, a few thousand steps later, one sees that the total energy plummets when plotting using \texttt{gnuplot}. Since the execution ended after the 2.1 millionth step, a continuation of this system with an extra 2.1 million steps was executed. Thus a total of 4.2 million MC steps had to be taken. This second time, the total energy remained low and constant, about 0.0006 eV lower, and an \texttt{aviz} plot displayed two silicon planes being perpendicular to each another.

\begin{figure}[htbp]
\centering
\subfigure[Systems with $T=0.00250$ eV and $T= 0.00275$ eV are tending toward order, while others remain at disorder.]
{\includegraphics[width=6.2cm]{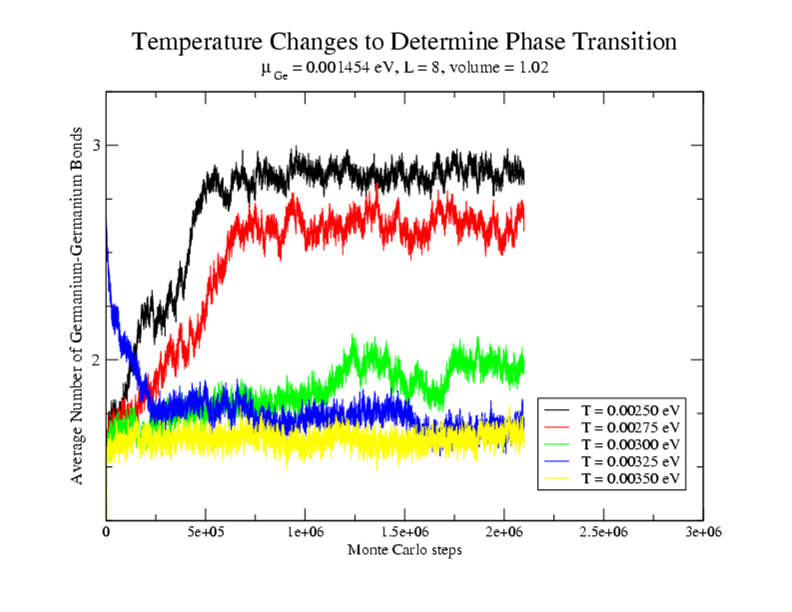}} \qquad 
\subfigure[Systems with $T=0.00390$ eV and $T= 0.00400$ eV are tending toward disorder, while others tend toward a stable, ordered system.]
{\includegraphics[width=6.2cm]{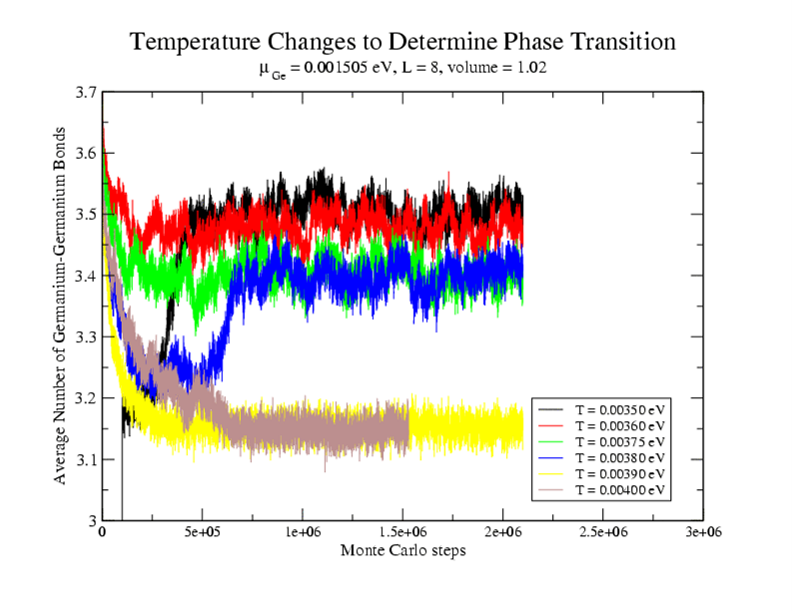}}
\label{fig:plot-01}
\end{figure}

\begin{figure}[htbp]
\centering
\subfigure 
{\includegraphics[width=6.2cm]{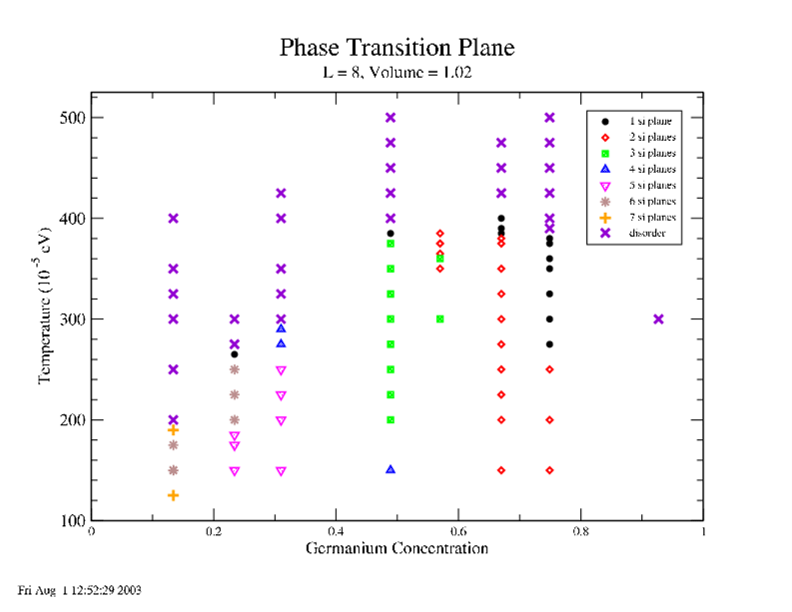}} \qquad 
\subfigure
{\includegraphics[width=6.2cm]{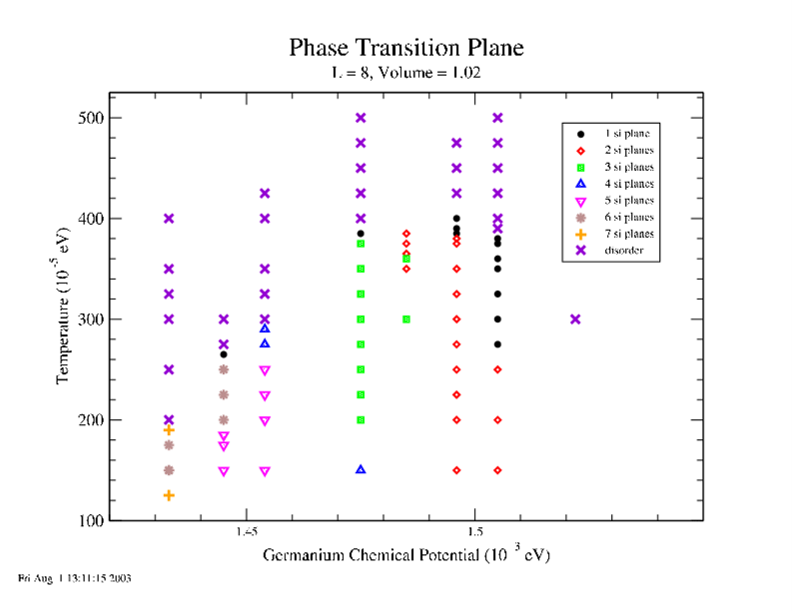}} 
\label{fig:plot-03}
\end{figure}

 \begin{figure}[htbp]
\centering
\subfigure
{\includegraphics[width=5.2cm]{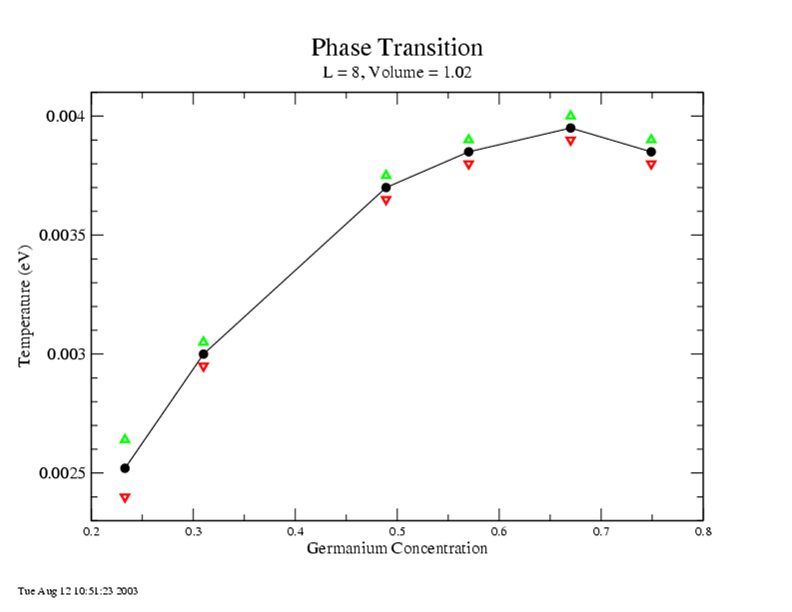}}\qquad 
\subfigure
{\includegraphics[width=5.2cm]{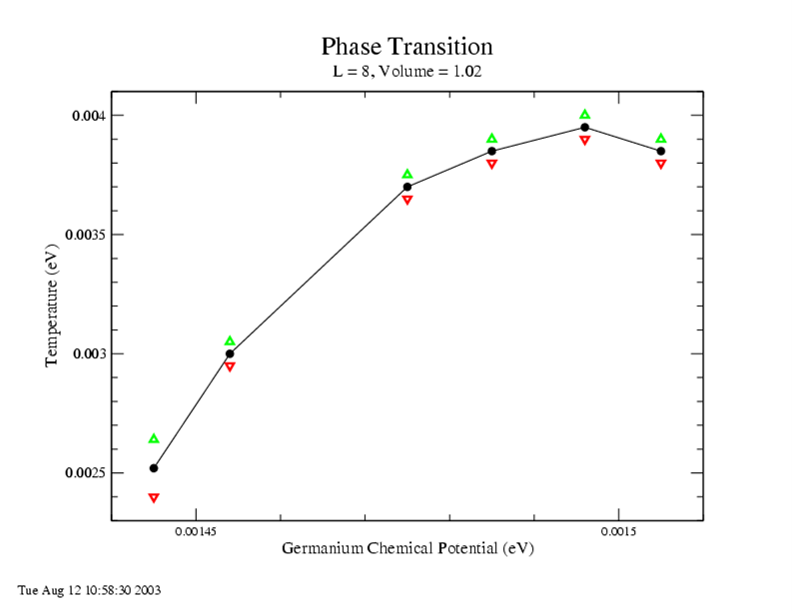}} 
\label{fig:plot-05}
\end{figure}

 \begin{figure}[htbp]
\centering
\subfigure[Ge chemical potential of 0.001496 eV and temperature
of 0.00275 eV. 
$L = 8$ and volume at 1.02. 
Total of 4.2 million MC steps are shown. Hysteresis observed.]
{\includegraphics[width=5.1cm]{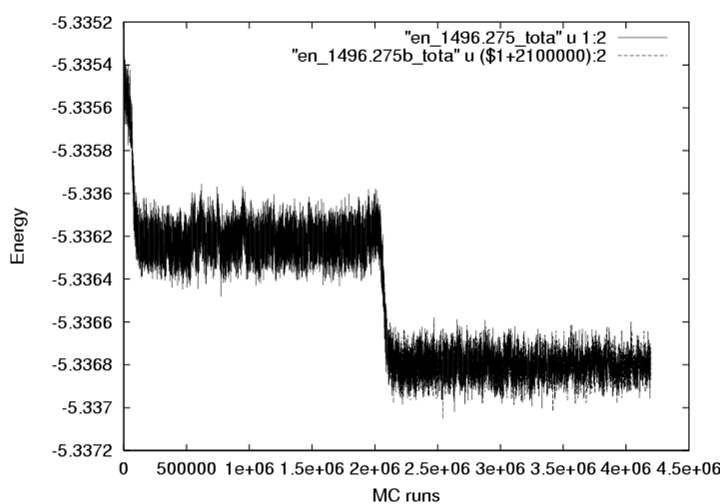}} \qquad 
\subfigure[Final equilibrium state. Snapshot of the $8$-system after 4.2 million MC steps. Different random number was used. Only the silicon atoms are shown. The system has germanium chemical potential of $0.001496$ eV, germanium concentration of $67\%$, and temperature at $0.00275$ eV. Its volume is at $1.02$.]
{\includegraphics[width=5.1cm]{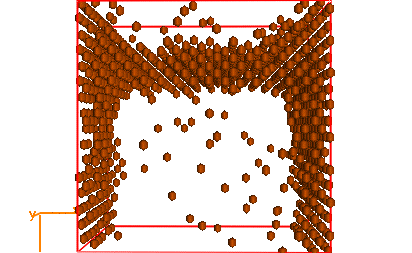}} \qquad 
\subfigure[Metastable state. Snapshot of the $8$-system after 1.4 million MC steps. Random number 121143179 was used. Shown only the silicon atoms. The system has germanium chemical potential of 0.001496 eV and temperature at 0.00275 eV. Its volume is 1.02.]
{\includegraphics[width=5.1cm]{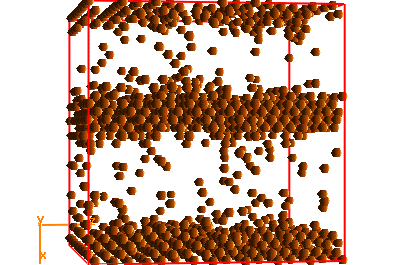}}
\label{fig:0.001496}
\end{figure}

Usually, when silicon planes form, they are mutually perpendicular to one another. But in this execution before the final several thousand steps, the snapshot of the two distinct silicon planes, both with similar thickness, were parallel to each another, instead of being mutually perpendicular like in all the other cases. Although both the metastable snapshot and the final shot showed two silicon planes clearly, it's obviously clear that two planes parallel to one another do not result in the lowest energy state. Thus, mutually perpendicular silicon planes give its final ordered state.

All random numbers must have nine digits, where the last digit must be an odd number in order for the program to be executed smoothly. The random number that was used for this particular system was 121143179. Using this random number, this was retested by shifting the temperature and changing the germanium chemical potential. In one case, a previously tested system that had two thick silicon planes has three equally thin silicon planes after using this random number. In another case, shorter computer time was required to reduce the system's energy and reach equilibrium compared to the previously tested systems with the same germanium chemical potential and temperature but with different random number. But in most cases, not many changes were detected. Relative thickness of silicon planes were similar, as well as the number of silicon planes, the number of germanium-to-germanium bonds, the length of time needed to reach equilibrium, and its equilibrium energy.

\begin{figure}[htbp]
\centering
\begin{minipage}[b]{.4\textwidth}
\includegraphics[width=6cm]{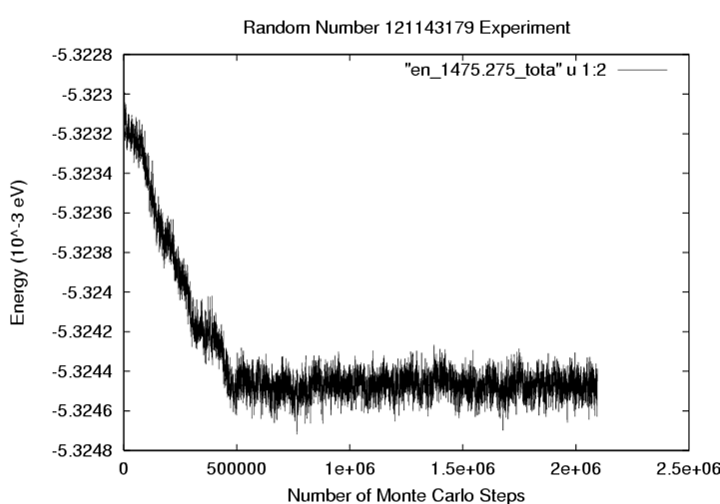}
\caption{ 
}
\label{fig:hysteresis-04}
\end{minipage}   \qquad 
\begin{minipage}[b]{.4\textwidth}
\includegraphics[width=6cm]{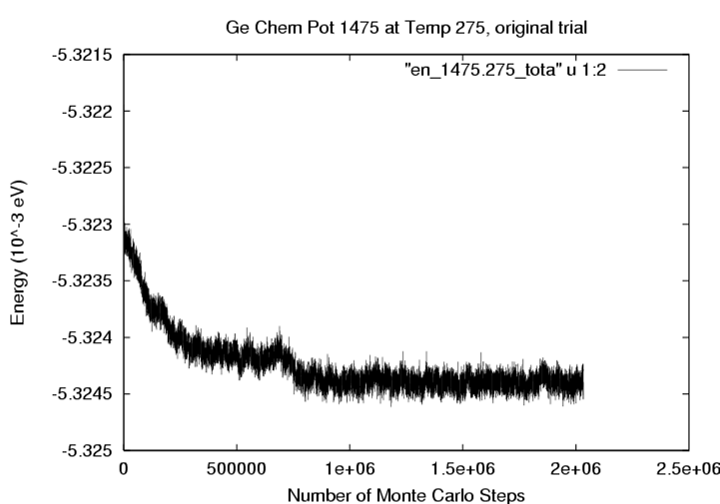}
\caption{ 
}
\label{fig:hysteresis-05}
\end{minipage}
\end{figure}

What makes this random number unique is not only by its output file, but when submitting a collection of input data, this random number is held fixed until the entire file has been executed. That is, since each run must be sufficiently long, the MC steps were separated into shorter executions and this file was programmed to automatically submit each of these separate input files after finishing the current run. When any other random number was chosen for an execution, the computer automatically changed its random number as each input file was consecutively and automatically submitted. But when this random number, one among other particular and unique random numbers, is part of the input files, the system executes all its input files using only this random number. 
 
Finally, hysteresis is affected as the germanium concentration changes. Holding the system's temperature constant, the number of germanium-to-germanium bonds or perhaps its total energy may depend on whether the germanium concentration is increasing or decreasing. The slowed down effect may affect the object in consideration if the system is used for technology. That is, a system with a greater gap in hysteresis will affect a machine when transporting or sending information. Also, when using the system for other uses, for chemistry or for health, hysteresis may affect medicine and the human body. The following plots give a better idea on hysteresis.

\begin{figure}[htbp]
\centering
\begin{minipage}[b]{.4\textwidth}
\includegraphics[width=6cm]{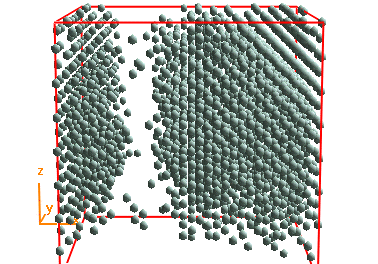}
\caption{This system was executed using random number 121143179, Ge chemical potential of 0.001475 eV and temperature at 0.00275 eV. The volume is 1.02 and $L = 8$. Shown are Ge. There are three silicon planes, each parallel to the rectangular coordinate axes. One silicon plane is not visible unless the system has been rotated at least $90^{\circ}$. 
}
\label{fig:hysteresis-06}
\end{minipage}\qquad 
\begin{minipage}[b]{.4\textwidth}
\includegraphics[width=6cm]{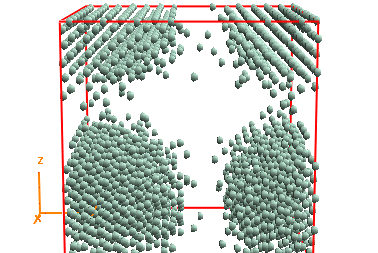}
\caption{This system was executed using another random number. Its Ge chemical potential is 0.001475 eV, and its temperature is 0.00275 eV. The volume is at 1.02 and $L = 8$. Shown are germaniums. There are two silicon planes, one along the $z$-axis and the other along the $y$-axis. 2.1 million MC steps were taken.
}
\label{fig:hysteresis-07}
\end{minipage}
\end{figure}

 \begin{figure}[htbp]
\centering
\subfigure
{\includegraphics[width=5.2cm]{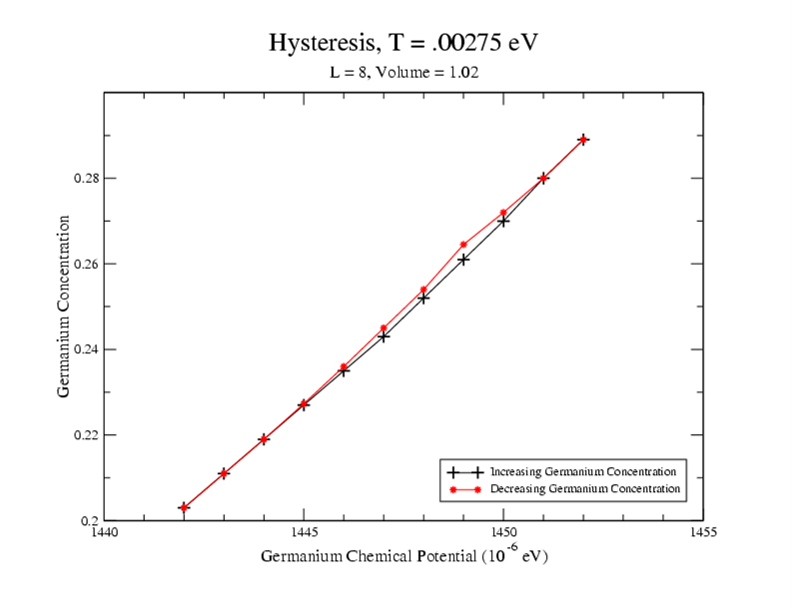}}\qquad 
\subfigure
{\includegraphics[width=5.2cm]{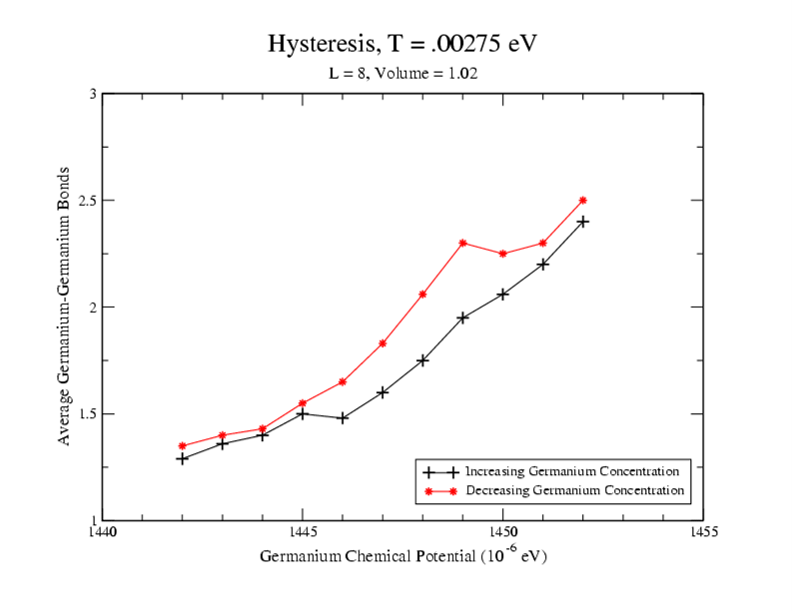}} \qquad 
\subfigure
{\includegraphics[width=5.2cm]{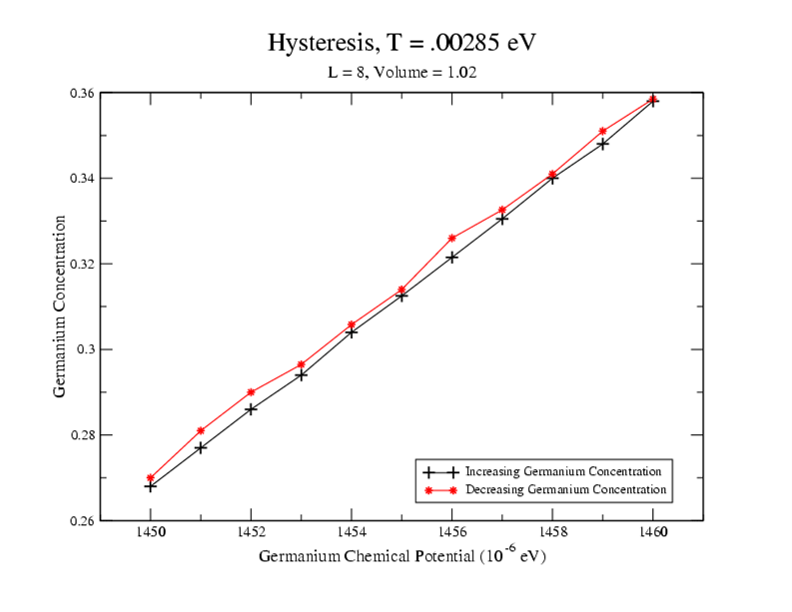}} 
\label{fig:hysteresis-08}
\end{figure}

\begin{figure}[htbp]
\centering
\subfigure
{\includegraphics[width=5.2cm]{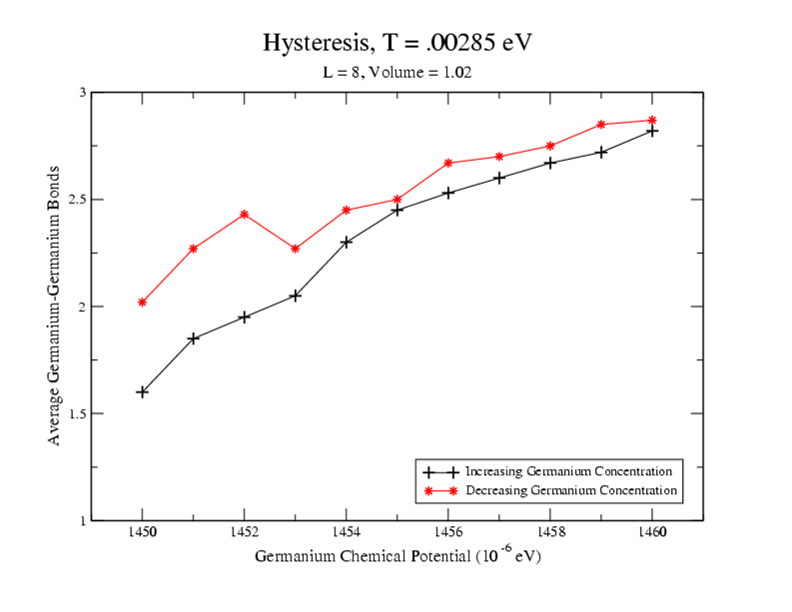}} \qquad 
\subfigure
{\includegraphics[width=5.2cm]{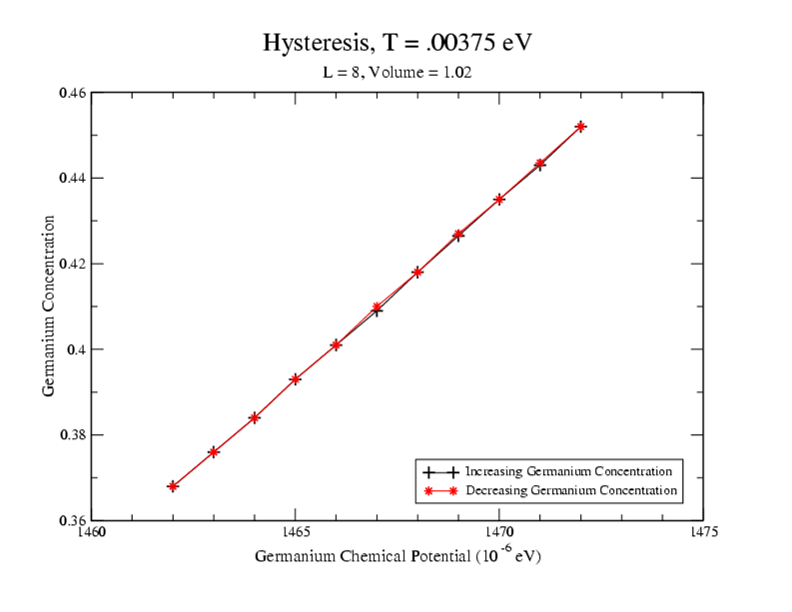}}\qquad 
\subfigure
{\includegraphics[width=5.2cm]{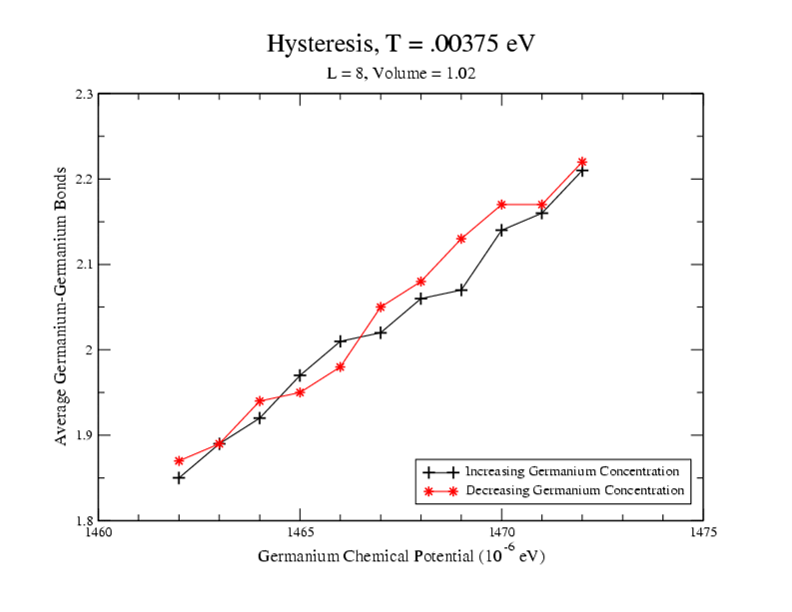}} 
\label{fig:hysteresis-10}
\end{figure}

\section{Discussion}
\label{section:discussion}

One problem that may occur in our simulation is that the MC runs must be sufficiently long enough. But how does one know that a system has been given sufficiently enough computation time? We may obtain another case, e.g., (o), (p), (q), with germanium chemical potential at 0.001496 eV (with about $66.7\%$ germanium) and its temperature at 0.00275 eV with a different random number, but with still unique and outputs an exceptional behavior in the atomic system. One way is to check if silicon planes or a symmetry can be seen. A better way to check is by comparing data. Do the measurements make sense with the rest of the output? If not, what other parameters need to be modified? Should the input be executed longer or is this a special behavior? 

Another mistake we might run across is putting in wrong input data or having a typographical error. An incorrect command can produce wrong data, interrupt a submitted run or not be executed at all. In addition, forgetting to modify the random number generator for each simulation will produce biased results. 

Finally, since time and memory are limited, we initiated this research by executing small systems and then expanding to larger atomic systems in order to obtain interesting results. With more memory and computation power, it may be possible to observe further instances of hysteresis in the Si-Ge system.

\subsection*{Acknowledgment}   
The author would like to thank David P. Landau for the opportunity to conduct research in summer of 2003, and she would also like to thank Francesca Tavazza for late night discussions. The author also thanks Mike Caplinger for installing Unix into her computer in order to run multiple simulations for this project; she was given access to considerable number of computational nodes so that she may obtain results for this manuscript. 
M.S.I. was partially supported by NSF \#0094422 through the Department of Physics and Astronomy at the University of Georgia.

\bibliography{si-ge}   

\def\cprime{$'$} \def\cprime{$'$} \def\cprime{$'$} \def\cprime{$'$}
\begin{thebibliography}{10}

\bibitem{binder1993monte}
Binder, K., Heermann, D., Roelofs, L., Mallinckrodt, A.~J., and McKay, S.,
  ``Monte {C}arlo simulation in statistical physics,'' {\em Computers in
  Physics}~{\bf 7}(2),  156--157 (1993).

\bibitem{landau2014guide}
Landau, D.~P. and Binder, K.,  [{\em {A guide to Monte Carlo simulations in
  statistical physics}}{\nolinebreak\hspace{0.1em}]}, Cambridge university
  press (2014).

\bibitem{hammersley2013monte}
Hammersley, J.,  [{\em Monte {C}arlo methods}{\nolinebreak\hspace{0.1em}]},
  Springer Science \& Business Media (2013).

\bibitem{hastings1970monte}
Hastings, W.~K., ``{Monte Carlo sampling methods using Markov chains and their
  applications},'' (1970).

\bibitem{kalos2009monte}
Kalos, M.~H. and Whitlock, P.~A.,  [{\em Monte {C}arlo
  methods}{\nolinebreak\hspace{0.1em}]}, John Wiley \& Sons (2009).

\bibitem{smith2013sequential}
Smith, A.,  [{\em {Sequential Monte Carlo methods in
  practice}}{\nolinebreak\hspace{0.1em}]}, Springer Science \& Business Media
  (2013).

\bibitem{heermann1990computer}
Heermann, D.~W., ``Computer-simulation methods,'' in [{\em Computer Simulation
  Methods in Theoretical Physics}{\nolinebreak\hspace{0.1em}]},   8--12,
  Springer (1990).

\bibitem{adler2002recent}
Adler, J., Hashibon, A., and Wagner, G., ``{Recent Developments in Computer
  Simulation Studies in Condensed Matter Physics, XIV},'' (2002).

\bibitem{conwell1952properties}
Conwell, E.~M., ``Properties of silicon and germanium,'' {\em Proceedings of
  the IRE}~{\bf 40}(11),  1327--1337 (1952).

\bibitem{conwell1958properties}
Conwell, E., ``Properties of silicon and germanium: {II},'' {\em Proceedings of
  the IRE}~{\bf 46}(6),  1281--1300 (1958).

\bibitem{shiraki2011silicon}
Shiraki, Y. and Usami, N.,  [{\em {Silicon-germanium (Si-Ge) nanostructures:
  Production, properties and applications in
  electronics}}{\nolinebreak\hspace{0.1em}]}, Elsevier (2011).

\bibitem{kelires1996microstructural}
Kelires, P.~C., ``Microstructural and elastic properties of
  silicon-germanium-carbon alloys,'' {\em Applied surface science}~{\bf 102},
  12--16 (1996).

\bibitem{lannoo1991atomic}
Lannoo, M. and Friedel, P., ``{Atomic and Electronic Structure of Surfaces,
  Springer Series in Surface Sciences Vol. 16},'' (1991).

\bibitem{lannoo2013atomic}
Lannoo, M. and Friedel, P.,  [{\em Atomic and electronic structure of surfaces:
  theoretical foundations}{\nolinebreak\hspace{0.1em}]}, vol.~16, Springer
  Science \& Business Media (2013).

\bibitem{Grace}
Grace, D.~M. and Karvonen, K., ``{Germanium-The Missing Element}.''
  \url{http://stopcancer.com/germaniumstor.htm} (2001).

\bibitem{Tavazza}
Tavazza, F.~M.,  [{\em {Investigation of bulk and surface properties of SI and
  SI-GE systems using Monte Carlo simulations and classical
  potentials}}{\nolinebreak\hspace{0.1em}]}, Department of Physics and
  Astronomy (2003).
\newblock Thesis (Ph.D.)--University of Georgia.

\bibitem{choi2002observation}
Choi, W., Chim, W., Heng, C., Teo, L., Ho, V., Ng, V., Antoniadis, D., and
  Fitzgerald, E., ``Observation of memory effect in germanium nanocrystals
  embedded in an amorphous silicon oxide matrix of a
  metal--insulator--semiconductor structure,'' {\em Applied Physics
  Letters}~{\bf 80}(11),  2014--2016 (2002).

\bibitem{hu1979superconducting}
Hu, E.~L. and Jackel, L.~D., ``Superconducting junctions utilizing a binary
  semiconductor barrier,'' (Mar.~20 1979).
\newblock US Patent 4,145,699.

\bibitem{pharr1992electrical}
Pharr, G., Oliver, W., Cook, R., Kirchner, P., Kroll, M., Dinger, T., and
  Clarke, D., ``Electrical resistance of metallic contacts on silicon and
  germanium during indentation,'' {\em Journal of materials research}~{\bf
  7}(4),  961--972 (1992).

\bibitem{yamazaki1975semiconductor}
Yamazaki, S. and Sugimura, Y., ``Semiconductor memories,'' (Apr.~15 1975).
\newblock US Patent 3,878,549.

\end{thebibliography}
\bibliographystyle{spiebib}

\end{document}